\documentstyle[a4wide,epsf,11pt,titlepage]{article}

\newcounter{nref}
\newcommand{\bbib}{%
  \renewcommand{\refname}{\large\bf References}%
  \begin{thebibliography}{99}%
  \setcounter{enumiv}{\thenref}}
\newcommand{\ebib}{%
  \end{thebibliography}%
  \setcounter{nref}{\arabic{enumiv}}}
\newcommand{\head}[3]{%
  \setcounter{nref}{0}%
  \thispagestyle{empty}%
  \section*{\LARGE\bf #1}%
  \stepcounter{section}%
  \addcontentsline{toc}{section}{#1}%
  \large\itshape%
  #2\\\vspace{0.1pt}\\%
  #3%
  \normalsize\upshape%
  \bigskip}

\newcommand{\sickamw}[1]{\langle#1\rangle_{\cal D}}
\newcommand{\sickawo}[1]{\langle#1\rangle_{{\cal D}_{0}}}

\begin{document}


\head{Backreaction in Cosmological Models}
{Christian Sicka$^1$, Thomas Buchert$\;^{2,1}$ and Martin Kerscher$^1$}
{$^1$Ludwig--Maximilians--Universit\"at, Theresienstra{\ss}e 37,
D--80333 M\"unchen, Germany\\
$^2$Theory Division, CERN, CH--1211 Gen\`eve 23, Switzerland\\ 
emails: sicka,buchert,kerscher@theorie.physik.uni-muenchen.de}

\subsection{The backreaction}

The  basic equations  for the  description of  structure  formation in
cosmology are Einstein's  laws for gravitationally interacting systems
or, with restrictions, Newton's law of gravity. Taking the Universe to
be filled with a  self--gravitating pressure--less fluid (`dust'), one
has to  deal with  a nonlinear system  of differential  equations.  To
simplify this  system of  differential equations one  commonly assumes
special symmetries.   Then, for  each set of  symmetries, we  obtain a
class of solutions.

For simplicity  and for some philosophical  reasons, most cosmological
models studied today are those  based on the assumption of homogeneity
and  isotropy.  Observationally  one can  find evidence  that supports
these assumptions on very large scales, the strongest being the almost
isotropy of the Cosmic  Microwave Background radiation after assigning
the whole  dipole to  our proper motion  relative to  this background.
However, on small and on intermediate scales up to several hundreds of
Mpcs,  there  are  strong  deviations from  homogeneity  and  isotropy
{}\cite{sicka.7}.
Here  the problem  arises  how  to relate  the  observations with  the
homogeneous and isotropic models.  The usual proposal for solving this
problem is to assume  that Friedmann--Lema\^\i tre models describe the
mean observables.
Such  mean  values  may  be  identified with  spatial  averages.   For
Newtonian fluid dynamics the averaging procedure has been discussed in
detail in  {}\cite{sicka.1}.  The key difference  between the averaged
models and  the standard  model can  be related to  the fact  that the
spatial average of a tensor field $\cal A$ over a domain comoving with
the fluid  (a ``Lagrangian  domain'') does not  commute with  the time
evolution of that tensor field:
\begin{eqnarray}
\label{sicka.eq1}
d_{t}\sickamw{{\cal A}}-\sickamw{d_{t}{\cal A}} =\sickamw{{\cal A}\theta} -
\sickamw{\theta}\sickamw{\cal A}   \ .
\end{eqnarray}
$\sickamw{\cal A}$  denotes the Euclidean spatial average  of $\cal A$
over a domain $\cal D$, and $\theta$ the local expansion rate.

By averaging Raychaudhuri's equation for a dust matter model and using
the  commutation  rule (\ref{sicka.eq1})  one  obtains a  differential
equation for the  averaged expansion rate.  This equation  can also be
written in a form similar  to the standard Friedmann equation, but now
with the domain dependent  scale factor $a_{\cal D}=V_{\cal D}^{1/3}$;
$V_{\cal  D}=|\cal D|$,  featuring an  additional  `backreaction' term
$\cal Q$:
\begin{equation}
\label{sicka.eq2}
3\frac{\ddot{a}_{\cal D}}{a_{\cal D}}+4\pi G\sickamw{\varrho}-\Lambda =
{\cal Q}\ \mbox{ with }\
{\cal Q}:=\frac{2}{3}(\sickamw{\theta^{2}}-\sickamw{\theta}^{2})
+2\sickamw{\omega^{2}-\sigma^{2}}\ .
\end{equation}
$\sigma$ denotes the  rate of shear and $\omega$  the rate of rotation
of an infinitesimal fluid element.

As  soon as inhomogeneities  are present  in the  domain, the  rate of
shear,  the rate  of  rotation and  the  expansion--rate are  nonzero.
Hence, ${\cal Q}\ne 0$ and the domain--dependent scale factor $a_{\cal
D}$ will behave in a different way compared with the scale factor of a
homogeneous--isotropic Friedmann cosmology.

\subsection{Spatially compact cosmologies without boundary}

The backreaction term can also  be written in the following form after
the application of Gau{\ss}' theorem:
\begin{equation}
{\cal Q}=\frac{1}{V_{\cal D}}\int_{\partial {\cal D}}
\left({\mathbf{u}}(\nabla\cdot{\mathbf{u}})-
({\mathbf{u}}\cdot\nabla){\mathbf{u}}\right)\cdot {\rm d}{\mathbf{S}}
-\frac{2}{3}\left(\frac{1}{V_{\cal D}}
\int_{\partial {\cal D}}{\mathbf{u}}\cdot {\rm d}{\mathbf{S}}\right)^{2} \ , 
\end{equation}
with the surface  $\partial{\cal D}$ bounding the domain  $\cal D$ and
the  surface-element  ${\rm  d}\mathbf{S}$; $\mathbf{u}$  denotes  the
peculiar velocity field.

If we  choose the compact  domain $\cal D$  to be the  whole universe,
then  backreaction  vanishes  for  all models  with  closed  3--spaces
(compact  without boundary).   For  example all  toroidal models  with
local   inhomogeneities  can  be   treated  globally   like  Friedmann
models. The homogeneous density in the Friedmann models is then simply
the density averaged over the  whole space.  From the same argument we
conclude that  no backreaction is present in  $N$--body simulations on
the periodicity scale.
However,  this  can  be  accomplished  only within  the  framework  of
Newtonian  theory.   In  General  Relativity  the  situation  is  more
involved  due to  the presence  of an  averaged contribution  from the
Ricci curvature, and the line  of arguments above is not conclusive in
that case {}\cite{sicka.2}.

\subsection{Effects of backreaction on intermediate scales}

On   intermediate  scales  the   Newtonian  `dust'   approximation  is
well--established  in   models  of  structure   formation.   With  the
inhomogeneous Friedmann  equation~(\ref{sicka.eq2}) we have  a tool to
quantify the time evolution of  the scale factor of spatial domains in
the Universe  and to relate  them to the global  background expansion.
Unfortunately, the dynamical evolution of the backreaction term is not
known.  Therefore,  we use the  Eulerian linear approximation  and the
`Zel'dovich approximation' to estimate  the effect of the backreaction
term. Such  a calculation  is only well--defined,  after all  what has
been  said, if  we assume  that the  Universe can  be  approximated by
spatially  flat  space  sections  and  that  the  inhomogeneities  are
subjected to periodic boundary conditions on some large scale. In this
way the  calculation of  the backreaction effect  on scales  below the
periodicity scale  can be investigated.  We want to emphasize  that no
conclusion about the global value of backreaction is possible, because
it vanishes  by assumption. This, of course,  restricts the generality
and shows the need to go to a general relativistic investigation. Russ
et al. {}\cite{sicka.6} have attempted this in a recent work. However,
their  assumptions  to start  with  are  too  restrictive, so  that  a
vanishing backreaction already follows from their basic equations (see
{}\cite{sicka.2} {}\cite{sicka.3}).

\subsubsection{Backreaction in Eulerian linear theory}

We use the linear approximation theory to calculate the time evolution
of the  velocity field. For an Einstein--de--Sitter  background and in
the initial stages of  structure formation the backreaction term $\cal
Q$ decreases {}\cite{sicka.3}:
\begin{eqnarray}
{\cal Q}^{\ell}=a^{-1}{\cal Q}_{0}\ \mbox{ with }\ 
a=\left(\frac{t}{t_{0}}\right)^{2/3}\ .
\end{eqnarray}
The dimensionless  contribution to the  expansion may be  estimated by
${\cal   Q}/(4\pi  G\varrho_H)$,   quantifying  the   impact   of  the
backreaction  $\cal  Q$  in   comparison  to  the  background  density
$\varrho_H$. This  dimensionless backreaction grows  proportionally to
the  variance of  the  density contrast  field $\delta^2_{\rm  r.m.s.}
\propto a^2$.  For  this type of linearization we  have to assume that
the deformation  of the comoving  volume is negligible.  We  can relax
this assumption if we work in the Lagrangian picture.  That is why the
`Zel'dovich approximation' is a better tool in this context.

\subsubsection{Backreaction in the `Zel'dovich approximation'}

Zel'dovich's approximation  is a subcase of solutions  to a Lagrangian
first--order  perturbation  approach {}\cite{sicka.4}{}\cite{sicka.5};
for the (dimensionless) trajectories of the fluid elements we have:
\begin{eqnarray}
f^{Z}({\mathbf{X}},t)=
a(t)({\mathbf{X}}+\xi (t)\nabla_{0}\psi({\mathbf{X}})) \ .
\end{eqnarray}
Here, $\nabla_{0}\psi({\mathbf{X}})$  denotes the initial displacement
field, $\xi(t)$  is a universal  function of time and  $\mathbf{X}$ is
the  initial  position  of   the  fluid  particles.   The  `Zel'dovich
approximation' is known to give a good picture for structure formation
also in the nonlinear regime until shell--crossing takes place.

It  is  useful to  calculate  the  scaled  backreaction term  $a_{\cal
D}^{6}{\cal  Q}$  in  the  `Zel'dovich approximation',  which  can  be
written in a compact form {}\cite{sicka.3}:
\begin{equation}
\label{sicka.eq3}
\begin{array}{rl}
a_{\cal D}^{6}{\cal Q}=
a^{6}\dot{\xi}^{2} & \left(2\sickawo{II_{0}}-
\frac{2}{3}\sickawo{I_{0}}^{2}+ \right.\\
& \ \left. +\xi(6\sickawo{III_{0}}
-\frac{2}{3}\sickawo{I_{0}}\sickawo{II_{0}}) 
+\xi^{2}(2\sickawo{I_{0}}\sickawo{III_{0}}
-\frac{2}{3}\sickawo{II_{0}}^{2})\right)\ ,
\end{array}
\end{equation} 
with  $\sickawo{I_{0}},\sickawo{II_{0}},\sickawo{III_{0}}$  being  the
first, the second and the  third scalar invariants of the tensor field
$\psi_{|ij}$  averaged  over   the  initial  domain  ${\cal  D}(t_0)$;
`$_{|}$'   denotes   the  derivation   with   respect  to   Lagrangian
coordinates.

\begin{figure}
\centerline{\epsfxsize=0.45\textwidth\epsffile{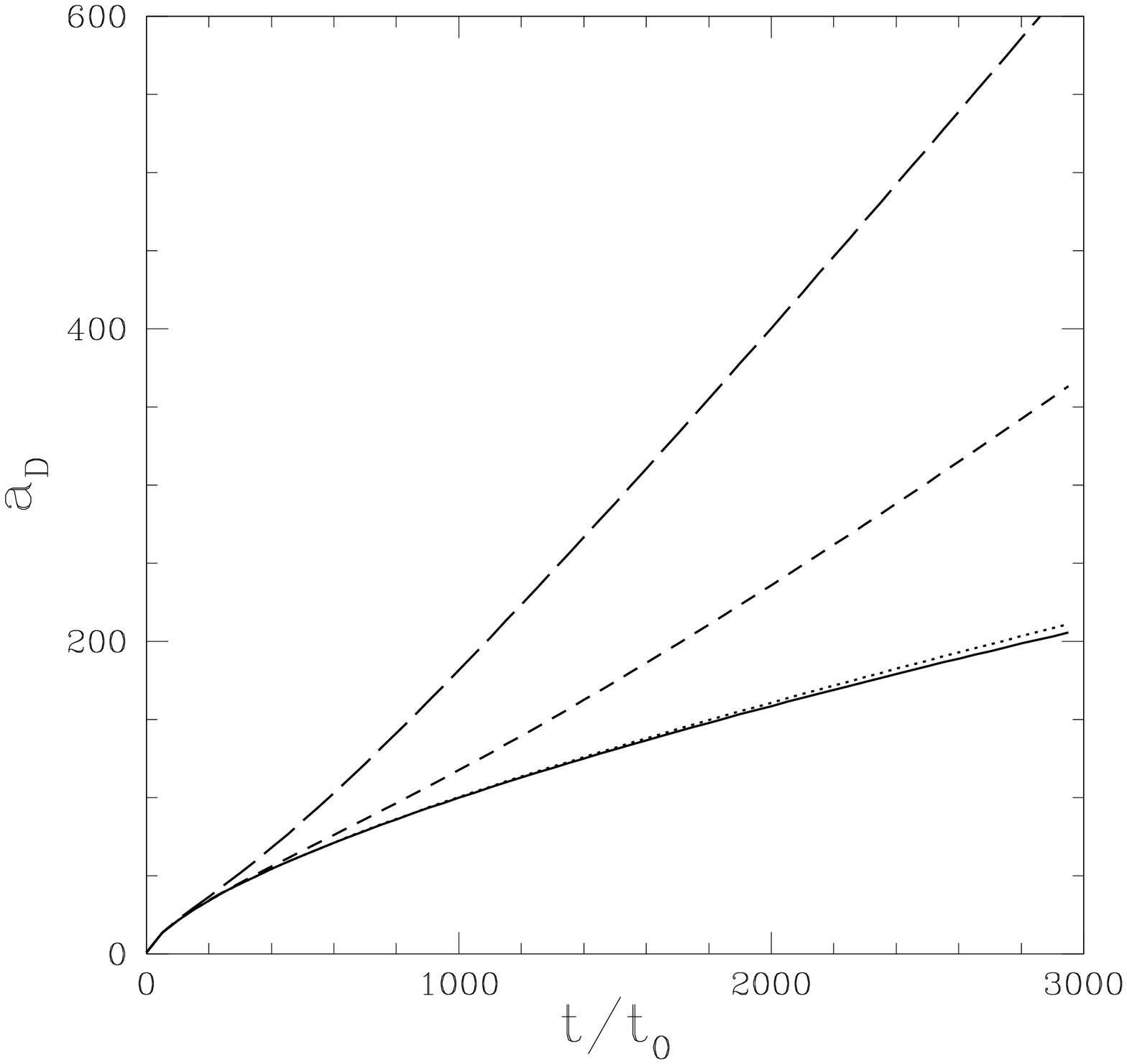}
            \epsfxsize=0.45\textwidth\epsffile{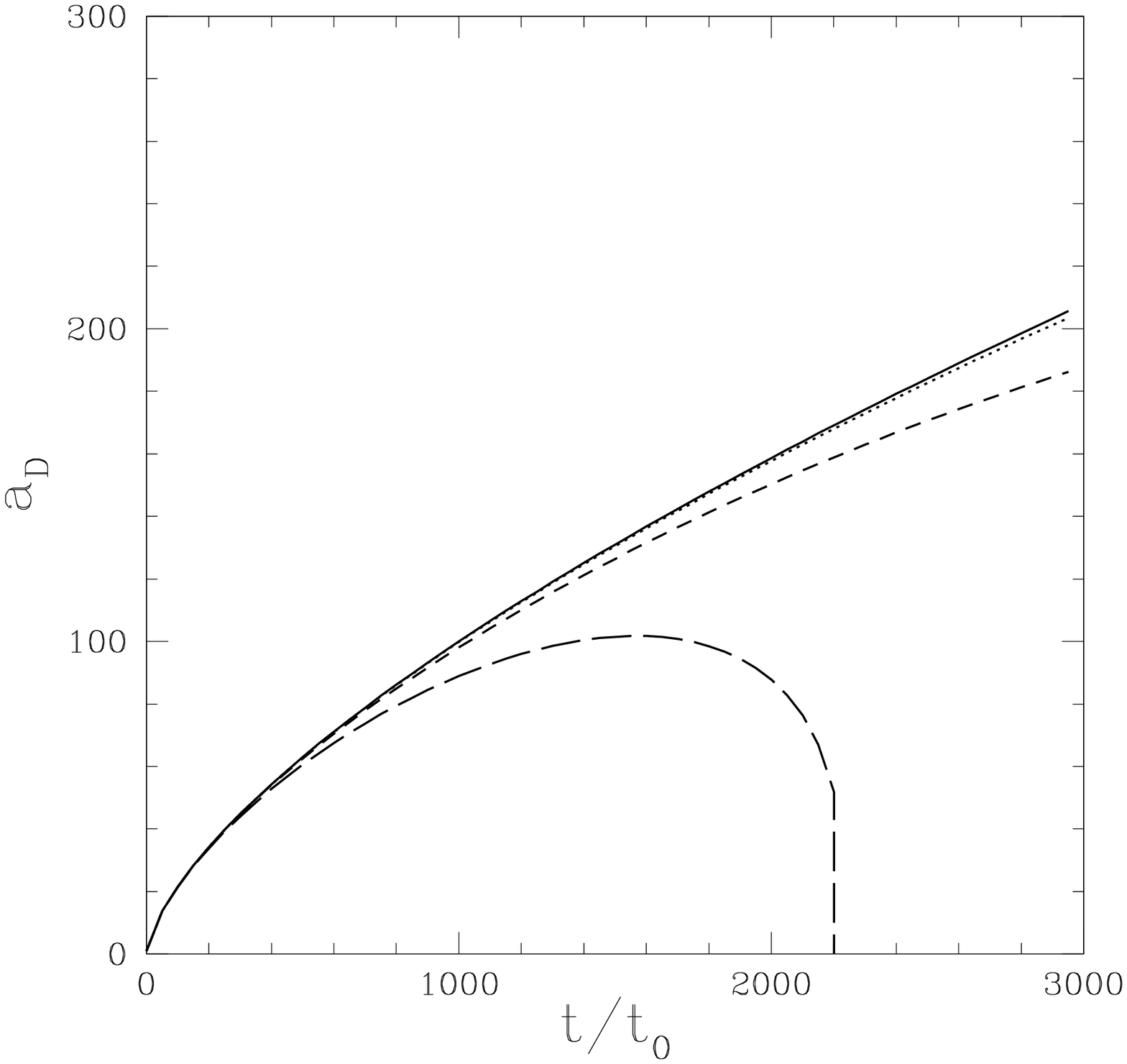}}
\caption{\label{sicka.fig1} In  the left  panel the time  evolution of
the  scale  factor  $a_{\cal  D}$ is  shown  for  $\sickawo{I_{0}}=0$,
$\sickawo{II_{0}}=10^{-6}$,  and $\sickawo{III_{0}}=10^{-8}$ (dotted),
$\sickawo{III_{0}}=10^{-6}$               (short              dashed),
$\sickawo{III_{0}}=10^{-5}$ (long dashed).  In the right panel we used
$\sickawo{I_{0}}=0$,          $\sickawo{II_{0}}=-10^{-6}$,         and
$\sickawo{III_{0}}=-10^{-8}$   (dotted),  $\sickawo{III_{0}}=-10^{-7}$
(short dashed), $\sickawo{III_{0}}=-10^{-6}$ (long dashed).  The solid
line  corresponds  to  an Einstein--de--Sitter  universe,  i.e.~${\cal
Q}=0$.}
\end{figure}

Using the  approximation (\ref{sicka.eq3}) for  the backreaction $\cal
Q$ we  solve the differential equation  (\ref{sicka.eq2}) for $a_{\cal
D}$ numerically.  The results  for various initial displacement fields
are  shown   in  Fig.~\ref{sicka.fig1}  for   an  Einstein--de--Sitter
background model with $a=(t/t_0)^{2/3}$.
We have  chosen $\sickawo{I_{0}}=0$ in both plots,  corresponding to a
mean  over--density  of  zero  in  the domain  ${\cal  D}_0$.   If  no
backreaction is  present, such domains should follow  the expansion of
the  background  model.    However,  with  $\sickawo{II_{0}}\ne0$  and
$\sickawo{III_{0}}\ne0$  we  have  ${\cal  Q}\ne0$  in  general.   The
accelerated  expansion visible in  the left  plot and  the accelerated
collapse in the  right plot is triggered only  by the small deviations
in the  initial displacement field as  described by $\sickawo{II_{0}}$
and $\sickawo{III_{0}}$.

Summarizing,  we  have   shown  that  the  backreaction  significantly
influences  the  dynamics of  Lagrangian  domains  in the  `Zel'dovich
approximation'.   Backreaction can invoke  a dynamics  that ressembles
cosmologies with a cosmological constant. Indeed every homogeneous and
isotropic  $\Lambda$--model can  be approximated  with  a backreaction
model  for suitably chosen  initial conditions.   Considering Gaussian
random  fields  we  will  estimate  the  effect  for  generic  initial
displacement fields {}\cite{sicka.3}.   Backreaction may, on the other
hand, provide us  with a new class of  collapse models that outperform
the  standard `top--hat'  model  (which excludes  backreaction by  its
restriction  to  spherical  symmetry).   Such models  may  deepen  our
understanding of the dynamics on cluster scales. Finally, backreaction
may substantiate the problem of missing mass: for domains dominated by
shear  fluctuations,  backreaction  can  act  as  a  ``dynamical  dark
matter'' component.

\subsection*{Acknowledgements}

This   work  is  supported   by  the   `Sonderforschungsbereich  f\"ur
Astroteilchenphysik SFB 375  der DFG'.  We thank  Claus  Beisbart for
valuable comments and discussions.



\bbib

\bibitem{sicka.5}
T.~Buchert,
\newblock Lagrangian theory of gravitational instability of
{F}riedmann--{L}ema{\^\i}tre cosmologies and the ``{Z}el'dovich 
approximation''
\newblock MNRAS {\bf 254}, 729, 1992.

\bibitem{sicka.1}
T.~Buchert and J.~Ehlers,
\newblock Averaging inhomogeneous {N}ewtonian cosmologies.
\newblock A\&A {\bf 320}, 1, 1997.

\bibitem{sicka.2}
T.~Buchert,
\newblock On average properties of inhomogeneous fluids in general
relativity I: dust cosmologies.
\newblock G.R.G. in press (gr-qc/9906015), 1999.

\bibitem{sicka.3}
T.~Buchert, M.~Kerscher and C.~Sicka,
\newblock Backreaction of inhomogeneities on the expansion: 
a dynamical approach to cosmic variance.
\newblock in preparation, 1999.

\bibitem{sicka.7}
M.~Kerscher, J.~Schmalzing, T.~Buchert and H.~Wagner,
\newblock Fluctuations in the IRAS 1.2 Jy Catalogue.
\newblock A\&A {\bf 333}, 1, 1998.

\bibitem{sicka.6}
H.~Russ, M.H.~Soffel, M.~Kasai and G.~B{\"o}rner,
\newblock Age of the Universe: influence of the inhomogeneities on
the global expansion-factor 
\newblock Phys.~Rev.~D {\bf 56}, 2044, 1997.

\bibitem{sicka.4}
Ya.B.~Zel'dovich ,
\newblock Fragmentation of a homogeneous medium under the
action of gravitation.
\newblock Astrophysics {\bf 6}, 164, 1970.

\ebib
\end{document}